# MEDICI: A simple to use synthetic social network data generator

David F. Nettleton[1], Sergio Nettleton[2], Marc Canal i Farriol[1]

[1]Universitat Pompeu Fabra, Barcelona, Catalunya, Spain, [2]Universitat Politècnica de Catalunya, Barcelona, Catalunya, Spain.

## Abstract

This paper describes an application, called *Medici*, designed to produce synthetic data for social network graphs, which can be used for analysis, hypothesis testing and application development by researchers and practitioners in the field. It builds on previous work by providing an integrated system, and a user friendly screen interface. It can be run with default values to produce graph data and statistics, which can then be used for further processing. The system is made publicly available in a Github Java project. The annex provides a user manual with a screen by screen guide.

## 1 - Introduction

The use of online social networks has been steadily increasing and evolving over the last decade, since they first became available to the general public in the 2000s. This has created a great interest for commercial and academic reasons in studying human behaviour in the online social network environment in particular and Internet in general. Behaviour rules have emerged on how users tend to group by affinities, how are they interconnected, which are the key demographics, how to capture and evaluate activity, information propagation, and the general dynamics of what makes social networks tick, all of which has made this a fascinating field of study [6]. On the other hand, privacy issues have raised concerns that major corporations use our data in a market where the control is lost to third parties. However, after the initial mercantilist focus, efforts to redirect research and applications for social good and taking into account ethical considerations are now becoming main-stream, especially with think-tank and government backing.

The algorithm for synthetic data generation was first developed by Nettleton in [7, 8], and in [9] Nettleton and Salas applied a preliminary version of the seed assignment approach to a data privacy application, in which the seeds were assigned to k-anonymous groups in order to anonymize them.

The *Medici* application described in this paper responds to some of the aforementioned issues, by offering a synthetic data generator for online social network graphs, which mitigates the need for accumulating real personal data of users, and can be useful for applied research and development in fields such as population studies for public resource assignment and medical care, pandemic data analysis, among others.

*Medici* addresses the need for an application which is accessible for general users with medium level skills, whereas it also can be used as a research support tool for more advanced users. The graphical user interface is designed to lower the entry barrier for the former type of users, and whose focus is on the data analysis *per se* rather than the process to obtain the data. A priority has been given to retain access to the functionality and control parameters, while maintaining a simple and user-friendly approach to allow the less process oriented users to obtain the data they need.

The system can be initially run in default mode with completely pre-assigned settings and example graph and community files. The user can go directly to the "generate data" tab and then to the "results" tab to see the statistics of the output data. This will enable the user to become familiar with the system, before progressing to modifying the profile to community assignments, the profile definitions, and finally the input graph and communities.

The paper is organized as follows: firstly, some theoretical background is given regarding the algorithms: **Rmat** (graph structure generator), **Louvain** (community labelling). This is described in relation to the functionality of the application and the sequence of steps for data processing. Next, a detailed description is given of the **Medici** algorithm and how it assigns and propagates user data through the graph structure. This is followed by the empirical testing section, first a description of the experimental setup, then examples of the data generated and benchmarking of deviations between different executions. The conclusions finalize the paper. In the annexes a full "user manual" is given (Annex A) and the pseudo-code of the data generator/propagator (Annex B). The full system source code and runtime is available at our public Github project [10].

**2- Background to third party algorithms**

The following describes the two "third party" algorithms: RMAT and Louvain, which generate the graph and label the communities, respectively. The third (propietary) algorithm, *Medici*, which assigns and propagates the data, is described later in Section 3.2.

**2.1 - Rmat**

Chakrabarti et al. [4] present R-Mat, a recursive model for graph mining. The objective of R-Mat is to model an existing graph of real data, thus deriving its parameterization in terms of given descriptor variables. A typical adjacency matrix of {0, 1} values is used to represent the graph (nodes, edges). The authors state that one of the challenges in modeling real graphs, such as social networks, is replicating the power law distributions, skew distributions, and other reported structures, such as the "bow-tie" and the "jellyfish", while maintaining a small diameter for the graph. The computation cost of generating the graph is also an issue. The authors indicate that a model of a social network must display a "community structure". For example, soccer and automobile enthusiasts, the latter of which can have sub-groups such as motorcycle and car enthusiasts. There are also cross-links between communities which denote persons with diverse interests (e.g. soccer *and* automobiles). In order to represent this, a recursive partitioning is carried out, which can be considered as a binomial cascade in two dimensions. The expected number of nodes $c_k$ with out-degree $k$ is given by:

$$c_k = \binom{E}{k} \sum_{i=0}^{n} \binom{n}{i} \left[ \rho^{n-i}(1-\rho)^i \right]^k \left[ 1 - \rho^{n-i}(1-\rho)^i \right]^{E-k} \qquad (1)$$

where $2^n$ is the number of nodes in the R-MAT graph (typically n = log$_2$N), ρ = probability of an edge falling into partition *a* + probability of an edge falling into partition *b*, and E is the number of edges in the real graph. The authors tested the method on two real datasets, "epinions" and "clickstream". Descriptive parameters are used such as degree distributions, number of reachable pairs, number of hops, effective diameter and stress distribution.

**2.2 - Louvain**

**The 'Louvain' method [2]:** this can be considered an optimization of Newman and Girvan's method [11], in terms of computational cost. Firstly, it looks for smaller communities by optimizing modularity locally. As a second step, it aggregates nodes of the same community and builds a new network whose nodes are the communities. These two steps are repeated iteratively until the modularity value is maximized. The optimization is based on evaluating the modularity gain, which is done by performing a local calculation of the change in modularity for a given community, caused by moving each node from it to an adjacent community. With each iteration the number of nodes to test quickly reduces (due to the aggregation of the corresponding nodes), and the computational cost is reduced in the same order.

**Modularity** [11]: During processing, the graph is successively divided in components, and the correctness of the community partitions is measured. The quality metric used for a given community is called the modularity. For a graph divided into *k* communities, a symmetrical matrix *e* is defined of order *k2* whose elements $e_{ij}$ are the subset of edges from the total graph which connect the nodes of communities i and j. The trace of matrix *e*, denoted as $Tr\, e = \sum_i e_{ii}$ gives the fraction of edges in the graph which connect nodes of the same community. Hence, a good division in communities should obtain a high value for the trace of matrix *e*. However, this value alone is not sufficient as a good quality indicator, given that if all the edges are placed in the same community this would give the maximum value $Tr\, e = 1$, but without having created any useful structure. Thus it is necessary to define the sum of the rows $a_i = \sum_j e_{ij}$, which represents the fraction of edges which connect nodes of community $i$. Following on from this, the modularity metric was defined as:

$$Q = \sum_i(e_{ii} - a_i^2) = Tr\ e - ||e^2|| \tag{2}$$

Where $||x||$ indicates the sum of the elements of matrix $x$. This parameter measures the fraction of edges in the graph which connect vertices in the same community, minus the expected value of the same number of edges in the graph with the same community partitions but with random connections between their respective nodes. If the number of intra-community edges shows no improvement on the expected value, then the modularity would be Q=0. On the other hand, Q approaches a maximum value of 1 when the community structure is strong. According to [11], the usual empirical range for Q is between 0.3 and 0.7. The modularity is calculated after each iteration of the elicitation algorithm, when two new components have been created due to the elimination of an edge. At this point a test is made to see if a global maximum, or some predefined expected maximum, has been reached.

For the current release of Medici, we have tweaked Louvain to produce exactly 10 communities (0 to 9), using the "resolution" as an optimization parameter (approximating to 1). Note that the "resolution" parameter was incorporated in the Gephi implementation of Louvain from the idea of Lambiotte et al. [5]. A resolution closer to 1 implies a better "quality" of the result, which is related to the "modularity" quality measure and the optimum partitioning. In our implementation, if the algorithm does not converge to 10 communities in the predefined timeout (30 seconds), we aggregate communities 9 to N in one community (labeled as 9). The community size after resolution optimization tends to fragment quickly for communities 9 to N, and the aggregate community (9) is generally one of the smallest with respect to communities 0 to 8 (less than 10% of the sum).

### 3– Medici

In the following, first an overall vision is provided in Section 3.1, then details of the data assignment and propagation algorithm are given in Section 3.2.

### 3.1- Overall vision

The following describes the user interface and functionality for the synthetic data generator for online social networks. It follows an easy to use "workflow" sequential structure guiding the user from the initial data entry starting on the leftmost tab of the main screen through the data definition and generation, followed by analysis of the generated data and export on the rightmost tab of the main screen.

Improvements in this version include the incorporation of the RMAT [4] and Louvain [2] algorithms to create the graph and assign the community labels, respectively. However, the user can provide their own graph and community assignment files if s/he wishes. Apart from the "static" information about each user, the system now also allows the assignment of dynamic data, such as simple "likes". This can serve as the basis so that the user can customize and develop more sophisticated simulations. For this, we make available the Java source code [10,3] in our Github repository.

Hence the file inputs to the program are the social network skeleton structure (nodes and links) where each node has been assigned a "community label". If the user wants to do this themselves it can be done manually or it can be done with an automated algorithm such as Louvain [2] (also available in Gephi [1] and included in MEDICI "as is").

One limitation is that there must be 10 communities and 10 profiles. Though this seems a bind, if you want less communities you can define some with very few members (by controlling the profile percentaje definitions) so they are residual and have a minimum effect on the overall network. If you want less profiles you can define some which are identical.

For the graph definition, we have included the default RMAT algorithm. If you want to supply a graph from outside the program you can do that but there its possible that the Louvain algorithm cannot establish 10 communities (0 to 9), depending on the size and complexity of the graph. Note the Louvain algorithm has been modified to aggregate communities 9 to N as community 9, with a timeout of 30 seconds on the process.

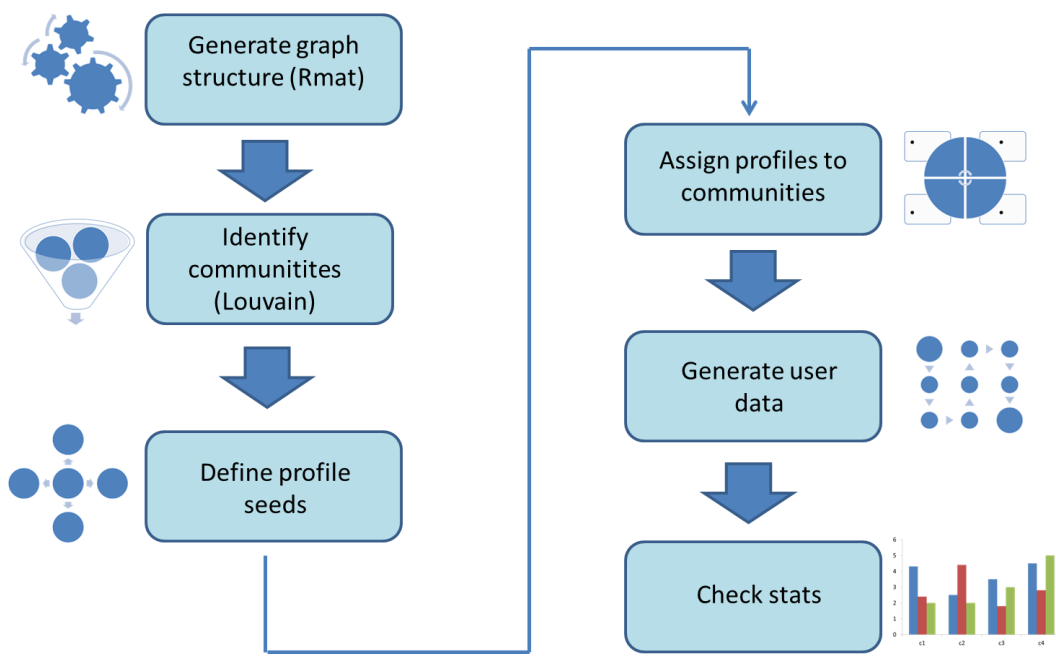

Fig 1: Processing sequence

**3.2- Data assignment and propagation**

The following gives a detailed description of the data assignment and propagation algorithm, the assignment of seeds, neighbors and any other nodes.

The **data generator** has the following four main steps:
      **(i)** Choose which nodes will be seeds in each community

**(ii)** Assign prototype profile to seeds in each community (profile *x* → community *y*)

**(iii)** Assign data to neighbors of each seed in function of seed profile (i.e. neighbors tend to be similar to seeds). This will have a similarity component (to neighbors with data assigned) and a random component (to promote a degree of diversity).

**(iv)** Assign data to remaining nodes (those which still have no data assigned from steps (ii) and (iii). This will have a similarity component (to neighbors with data assigned) and a random component (to promote a degree of diversity).

**Seed assignment**

The seed assignment has to comply with two criteria:

(i) Each seed node must be at least at distance 3 from any other seed node (in a given community), so that each sub-graph (with the seed as center) can be modified without perturbing the adjacent sub-graphs. In order to achieve this, seed nodes are added one by one, and checked for distance using a proximity routine. This routine checks if a given node is an immediate neighbor (distance = 1) or neighbor of a neighbor (distance = 2). If the given node is not a neighbor or a neighbor of neighbor, then it must be at distance 3 or more. The number of seed nodes assigned is initially defined as the total number of vertices in the graph divided by the average degree.

(ii) The distributions of the characteristics of the sub-graphs formed by the seed node neighborhoods must be within δ of the corresponding distributions in the complete graph. The characteristics are defined as: *degree distribution, clustering coefficient, distributions of attribute-values.*

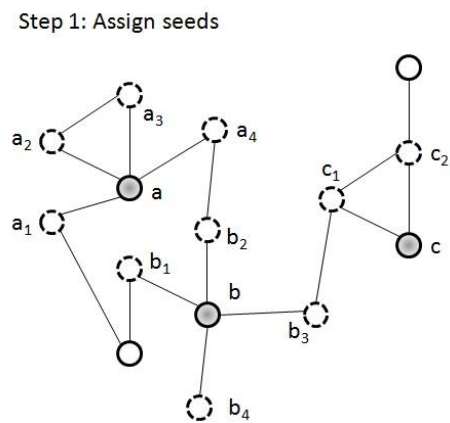

Fig 2: Seed assignment.

Fig. 2 shows a generic example of the seed assignment process, for which the seeds are labeled as *a*, *b* and *c*. We note that each seed is at a shortest distance of three from any other seed, hence the immediate neighborhoods do not overlap.

The assignment of the seed nodes is actually an optimization process. It is possible that the random assignment of seeds, especially the first seed, can result in a sub-optimal assignment. For example, if the first seed is assigned to a major hub node, the average path length to it of many nodes in the graph will be short.

Recall that we denote the number of seeds as $\sigma$ and we take as an upper-bound for the number of seeds to be assigned, the number $|V|/d_{AVG}$.

The seed assignment process has $T$ tries at randomly assigning $y$ see nodes. If, after $I$ iterations, $\sigma$ seeds have not been assigned, then $\sigma$ is reduced by one and another try is made. $\sigma$ is progressively reduced by one until $\sigma$ seeds are assigned. Finally, the process tries to add more seeds to the current configuration (to avoid the assignment of a sub-optimal number of seeds).

**Coverage:** the assignment of the seed nodes in the manner described has a coverage of between 20% and 50% of the nodes in the graph. This is because isolated nodes tend to remain between sub-graphs which cannot be assigned due to the minimum distance requirement between seeds (≥ 3) in the same community, and because the seed sub-graphs cannot overlap.

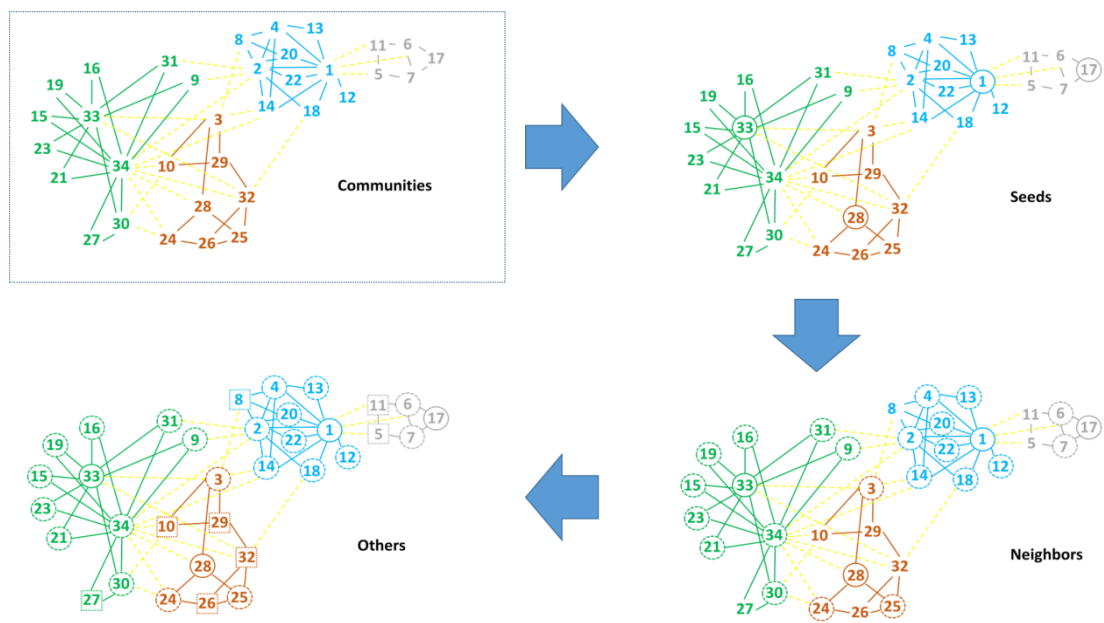

Fig. 3: Example of data propagation steps (Karate graph dataset): (Communities) four communities are identified; (Seeds) seeds assigned to each community; (Neighbors) neighbors of seeds are assigned; (Others) any remaining nodes are assigned.

Fig. 3 illustrates the data propagation mechanism. Fig. 3 "Communities" shows four communities identified in the Karate graph which will have been assigned by the Louvain algorithm, as indicated by the green, red, blue and grey colors. Next, in Fig. 3 "Seeds" the seeds are assigned to each community, as indicated by the circled nodes 33, 28, 1 and 17, respectively. In this example, due to the small size of the graph, only four seeds are used, one per community. At this point the profiles corresponding to each community are assigned to the seeds. This is followed by the assignment of the data to the neighbors of each seed (Fig. 3 "Neighbors") as indicated by the dashed circles. The neighbors are assigned values which "similar" to their seed profiles, depending on the "similarity theshold" defined. Finally, as shown in Fig. 3 "Others", any remaining nodes are assigned values which are either similar to the neighbors or randomly assigned, dependening on the control parameter values defined. These are indicated as dashed squares.

The number of seeds assigned guarantees a good "coverage" of the data, and tries to minimize the number of unassigned nodes (which are neither seeds nor neighbors of seeds). So the user can see the coverage and try different number of seeds to see which gives the best. By default, the number of seeds is equal to about 11% of the number of nodes in the graph. Also, seed nodes are chosen based on the HITS statistic, which indicates the nodes which are best connected in a community (i.e. those which have the most neighbors).

Thus, two aspects are optimized during the process:

(i) The maximum number of possible seeds is assigned, resulting in an optimal or close to optimal coverage of the complete graph, given the restriction that seeds should be at a distance of three or more from each other.
(ii) The distributions of the characteristics of the seed sub-graphs (seeds and their neighbors) has a given similarity to the corresponding distributions of the characteristics in the complete graph (all the nodes of the graph).

**4- Empirical Testing**

This section first describes the hardware and software used, together with performance considerations (4.1), followed by examples of the data generated by the system (4.2), and finally benchmarking of the deviation of the data distributions between successive executions of the system (4.3).

**4.1 - Setup, computational cost and performance**

The hardware used is a Lenovo laptop with Intel Core i7-7500U, 2.9Ghz, 2 processors, 12Gb RAM and Windows 10, 64 bits. The software used is Eclipse Java IDE for Web Developers Version 4.9.0 and Java Fx.

The Rmat algorithm is highly efficient and generates the 1k, 10k and 50k graphs almost instantly on the hardware and software indicated. The Louvain algorithm is less efficient and for the 1k and 10k graphs it labels the communities in less than 30 seconds. For the 50k graph a 30 second time limit was assigned, which labels 10 communities with a high quality modularity. The seed assignment is a computationally expensive process which takes less than 60 seconds for the 1k and 10k graphs (11% seeds) and less than 300 seconds for the 50k graph (15% seeds). For graphs up to 1M nodes it would be recommended to run the system on higher range hardware and for bigger graphs some of the system would need to be reprogrammed for parallel computation in a cloud environment.

Medici: A simple to use synthetic social network data generator                                    9## 4.2 - Example data generated

In Tables 1a and 1b are shown the data generated from one of the executions of the data generation for the 50k by 250k graph. The three rightmost columns give the number of users (or nodes) assigned to each attribute-value for (i) the whole graph, (ii) community 4 and (iii) community 1.

In order to interpret the results, we refer to the overall percentages assigned for each attribute value, and the profiles assigned to each community. For example, in the overall graph the users are biased towards young people (age 18-25 and 26-35), with somewhat more females than males, residence is fairly equitative, and so on.

In Community 4, there is a bias towards older people (age 66-75 and 76-85), mainly female, who live in San Jose, mainly Jewish, and so on. This is because the seed Profile 5 assigned to Community 4 has these characteristics. Likewise, in Community 1, there is a bias towards younger people (as in the whole graph), slightly more males, who live in Palo Alto, mainly Christian, and so on. This is because the seed Profile 2 assigned to Community 1 has these characteristics.

By comparing the results with the overall attribute-value proportions for the whole graph and the seed profiles assigned to each community, we have validated that the output produced is as expected, taking into account the stochastic factor in the assignment which will cause some flunctuations.

Table 1a: Attribute-value assignments (age to marital) - 50k nodes by 250k edges.

|           |              | ALL       | COMMUNITY 4 | COMMUNITY 1 |
|-----------|--------------|-----------|-------------|-------------|
| Attribute | Value        | Frequency | Frequency   | Frequency   |
| Age       | 18-25        | 10065     | 43          | 1935        |
|           | 26-35        | 10247     | 48          | 706         |
|           | 36-45        | 7517      | 19          | 173         |
|           | 46-55        | 2620      | 17          | 87          |
|           | 56-65        | 8726      | 92          | 71          |
|           | 66-75        | 2922      | 841         | 60          |
|           | 76-85        | 1136      | 107         | 88          |
| Gender    | Male         | 17041     | 101         | 1904        |
|           | Female       | 26457     | 1067        | 1220        |
| Residence | PaloAlto     | 5040      | 105         | 1989        |
|           | SantaBarbara | 9541      | 109         | 350         |
|           | Winthrop     | 8123      | 50          | 141         |
|           | Boston       | 8371      | 31          | 162         |
|           | Cambridge    | 8032      | 41          | 176         |
|           | SanJose      | 4097      | 831         | 302         |
| Religion  | Buddhist     | 8617      | 16          | 115         |
|           | Christian    | 9056      | 280         | 2242        |
|           | Hindu        | 10962     | 13          | 137         |
|           | Jewish       | 4007      | 794         | 222         |

Nettleton, Nettleton and Canal i Farriol                                    Jan. 2021

| | | | | |
|---|---|---:|---:|---:|
| | Muslim | 3861 | 46 | 266 |
| | Sikh | 749 | 0 | 0 |
| | TraditionalSpirituality | 0 | 0 | 0 |
| | OtherReligion | 0 | 0 | 0 |
| | NoReligiousAffiliation | 4743 | 2 | 32 |
| **Marital** | Single | 14720 | 190 | 1839 |
| | Married | 14596 | 815 | 742 |
| | Divorced | 6043 | 78 | 285 |
| | Widowed | 7860 | 84 | 255 |

Table 1b: Attribute-value assignments(profession to like 3) - 50k nodes by 250k edges.

| | | ALL | COMMUNITY 4 | COMMUNITY 1 |
|---|---|---:|---:|---:|
| **Attribute** | **Value** | **Frequency** | **Frequency** | **Frequency** |
| **Profession** | Manager | 2998 | 13 | 167 |
| | Professional | 8142 | 35 | 248 |
| | Service | 2683 | 17 | 184 |
| | Salesandoffice | 12662 | 29 | 302 |
| | NaturalResourcesConstructionAndMaintenance | 0 | 0 | 0 |
| | ProductionTransportationAndMaterialMoving | 0 | 0 | 0 |
| | Student | 4368 | 45 | 1966 |
| **Political** | FarLeft | 3884 | 829 | 172 |
| | Left | 8718 | 185 | 377 |
| | CenterLeft | 5883 | 42 | 1935 |
| | Center | 12946 | 52 | 257 |
| | CenterRight | 7456 | 30 | 173 |
| | Right | 4118 | 24 | 182 |
| | FarRight | 198 | 5 | 15 |
| **Sexuality** | Asexual | 1120 | 29 | 70 |
| | Bisexual | 2162 | 44 | 178 |
| | Heterosexual | 38106 | 1049 | 2712 |
| | Homosexual | 2148 | 45 | 170 |
| **Like 1** | Entertainment | 15006 | 875 | 570 |
| | Music Artist | 7922 | 112 | 439 |
| | Drink Brand | 6116 | 80 | 1797 |
| | TV Show | 14211 | 101 | 309 |
| **Like 2** | Entertainment | 13192 | 156 | 567 |
| | Music Artist | 13043 | 112 | 435 |
| | TV Show | 10733 | 798 | 338 |
| | Drink Brand | 6261 | 101 | 1776 |
| **Like 3** | Music Artist | 6091 | 111 | 409 |
| | Entertainment | 10037 | 825 | 1874 |
| | Drink Brand | 13749 | 101 | 434 |
| | Soccer Club | 13337 | 130 | 400 |

**4.3 Deviation between successive executions**

The system has been initially tested by 3 fold validation for three graph dimensions (nodes, edges): 1k by 10k (Table 2), 10k by 100k (Table 3) and 50k by 250k (Table 4). The deviation for successive executions is detailed for each overall graph and for two selected community profile assignments. The results show a good stability, taking into account the stochastic nature of the data generator. In Table 2, for example, shows the deviations between three succesive runs for populating the same graph and community structure with 1k nodes and 10k edges. It can be seen that for the whole graph, the average and standard deviation is in general below 4%, with the exception of gender which is slightly higher. For two example communities, 1 and 4, the average and standard deviation does not pass 10%, with the exception of Gender (Community 1) and Like1 (Community 4). It is noted that, due to the stochastic nature of the see profile assignment and propagation, variations are expected, but the general distributions should be fairly consistent for the same graph, communities and seed profile assignments.

Table 2: 1k nodes by 10k edges.

| | All - deviation | | Community 1 - deviation | | Community 4 – deviation | |
|---|---|---|---|---|---|---|
| | **Avg.** | **Stdev.** | **Avg.** | **Stdev.** | **Avg.** | **Stdev.** |
| Age | 1.8 | 1.2 | 6.4 | 4.1 | 5.3 | 3.2 |
| Gender | 6.7 | 4.5 | 18.4 | 10.2 | 9.2 | 7.2 |
| Residence | 1.5 | 0.9 | 4.4 | 3.1 | 5.1 | 3.7 |
| Religion | 1.8 | 1.1 | 2.7 | 2.0 | 2.3 | 1.3 |
| Marital | 3.9 | 2.2 | 5.6 | 3.1 | 9.2 | 7.1 |
| Profession | 0.8 | 0.6 | 6.7 | 4.4 | 2.0 | 1.7 |
| Political | 1.5 | 0.9 | 4.9 | 3.2 | 3.5 | 2.0 |
| Sexuality | 1.7 | 0.9 | 4.6 | 3.0 | 6.6 | 3.6 |
| Like1 | 2.8 | 1.6 | 9.2 | 6.1 | 10.2 | 8.2 |
| Like2 | 2.1 | 1.6 | 9.7 | 6.1 | 9.2 | 5.1 |
| Like3 | 3.3 | 1.8 | 6.6 | 3.6 | 7.1 | 4.1 |

Tables 3 and 4 show similar statistics for scale up to progressively bigger graphs, and it can be seem that the deviations become progressively smaller for scale-up: the maximum average deviation for the 10k graph is 2.6% and for the 50k it is 1.5%

Table 3: 10k nodes by 100k edges.

| | All - deviation | | Community 1 - deviation | | Community 4 – deviation | |
|---|---|---|---|---|---|---|
| | **Avg.** | **Stdev.** | **Avg.** | **Stdev.** | **Avg.** | **Stdev.** |
| Age | 0.4 | 0.3 | 0.8 | 0.4 | 0.9 | 0.6 |
| Gender | 1.1 | 0.9 | 2.6 | 1.7 | 0.2 | 0.1 |
| Residence | 0.6 | 0.3 | 0.8 | 0.5 | 1.4 | 0.9 |
| Religion | 0.3 | 0.2 | 0.5 | 0.3 | 0.4 | 0.2 |
| Marital | 0.7 | 0.4 | 0.4 | 0.3 | 1.3 | 0.8 |
| Profession | 0.2 | 0.2 | 0.7 | 0.4 | 0.4 | 0.2 |
| Political | 0.5 | 0.3 | 0.7 | 0.4 | 0.7 | 0.4 |
| Sexuality | 0.5 | 0.3 | 1.0 | 0.6 | 0.5 | 0.3 |
| Like1 | 0.7 | 0.4 | 1.9 | 1.0 | 0.5 | 0.3 |
| Like2 | 1.1 | 0.6 | 1.9 | 1.0 | 1.4 | 0.8 |
| Like3 | 0.6 | 0.4 | 1.2 | 0.7 | 1.0 | 0.5 |

Table 4: 50k nodes by 250k edges.

| | All - deviation | | Community 1 - deviation | | Community 4 – deviation | |
|---|---|---|---|---|---|---|
| | Avg. | Stdev. | Avg. | Stdev. | Avg. | Stdev. |
| Age | 0.3 | 0.1 | 0.7 | 0.4 | 0.7 | 0.4 |
| Gender | 0.3 | 0.2 | 0.9 | 0.6 | 0.7 | 0.5 |
| Residence | 0.3 | 0.2 | 0.9 | 0.5 | 0.7 | 0.5 |
| Religion | 0.2 | 0.1 | 0.4 | 0.2 | 0.5 | 0.3 |
| Marital | 0.4 | 0.2 | 1.1 | 0.8 | 0.6 | 0.4 |
| Profession | 0.2 | 0.1 | 0.4 | 0.3 | 0.6 | 0.3 |
| Political | 0.2 | 0.1 | 0.6 | 0.4 | 0.9 | 0.5 |
| Sexuality | 0.3 | 0.1 | 1.1 | 0.6 | 0.7 | 0.5 |
| Like1 | 0.5 | 0.3 | 1.3 | 0.6 | 1.1 | 0.7 |
| Like2 | 0.4 | 0.3 | 1.5 | 0.9 | 0.6 | 0.4 |
| Like3 | 0.7 | 0.4 | 1.3 | 0.7 | 1.1 | 0.5 |

**5 - Summary**

The Medici application embodies a self-contained tool for generating synthetic data for small to medium size social network graphs. Using the Rmat and Louvain algorithms, the user can create a network from scratch and label the communities. Then, the user can define the seed profiles based on user demographics and likes, assign the seeds to the communities, and finally generate the data using the Medici algorithm.

Although the system currently has the stated limitations, the data diversity and volume offered can serve for many useful studies and applications. It will also serve for software developers to scale up the data processing and add more flexible functionalities, such as a variable number of communities and adding new attributes and their values. As future work, dynamic user activity data is planned to be incorporated into the user affinities with time and place information.

**Author contributions**
DN was responsible for developing the back end algorithms, improving the front-end interface/middleware, coordinating the software development, and writing of the paper. MC was responsible for implementing the initial version of the user screen interface and middleware. SN was responsible for improving the user screen interface and middleware, and integrating/testing the Rmat and Louvain algorithms in the system.

**Source code and system runtime**
The source code written in Java/Java FX and system runtime is publicly available in the indicated Github link [10]. *The system requires Eclipse/Java runtime (or similar) to be installed in your computer.*

**Annex A. Program Structure and User Guide**

In the following, the user interface screens are shown with an explanation of each, with a mention of any related functionality between the different screens.

The source code written in Java/Java FX and system runtime is publicly available in the Github link "https://github.com/dnettlet/MEDICI". *The system requires Eclipse/Java runtime (or similar) to be installed in your computer.*

As a first test run, it is recommended the user runs the system in default mode with completely pre-assigned settings and example graph and community files. Take a look at where the input, default and output files are located with the file explorer. Then go directly to the "generate data" tab, click the blue colored "generate data" button when it completes go to the "results" tab to see the statistics of the output data. This will enable the user to become familiar with the system, before progressing to modifying the profile to community assignments, the profile definitions, and finally the input graph and communities.

As a general vision, Figure 1 shows the dependencies between different screens – while the majority are independent (they are managed directly by rootLayout), some have inter-dependencies. For example, the attribute edit screen depends on the attribute screen, and "Prof. Comm Assign" depends on "Comm. & Profiles".

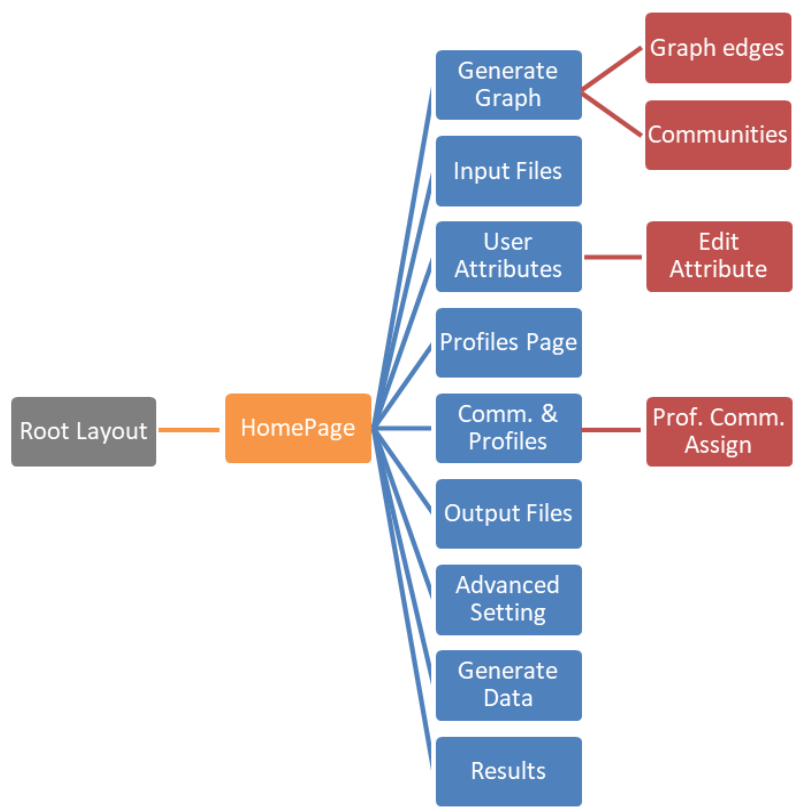

*Fig. 1. Hierarchy of interface screens*

Medici: A simple to use synthetic social network data generator 16**A.1 Root Layout**

The Root Layout is not a screen as such, but the basis for all the user interface screens, and contains just a bar in the upper menu. It is always visible given that it allows fast access to key functions.

It has a basic structure with different pull-down menus which offer the different options (see Figs. 2). The first menu is "File" (Fig. 2a), which gives access to the functionality of "Import Config" to import a configuration from a file; "Export Config" to export the actual configuration into a file; "Close" to finalize the execution and close the program.

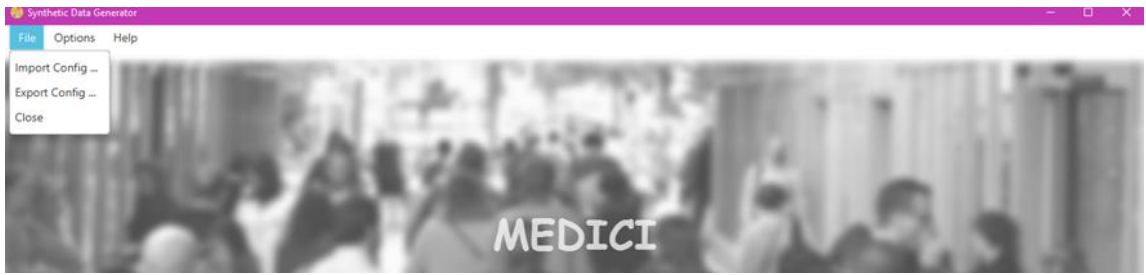

*Fig. 2a Main menu pulldown list (File)*

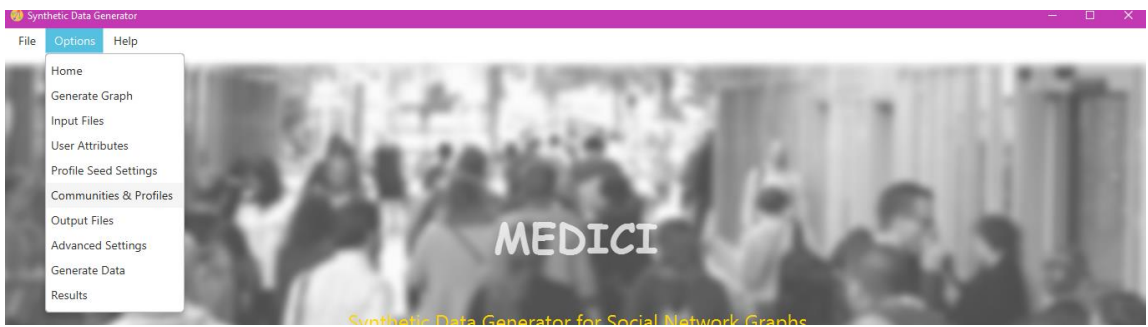

*Fig. 2b Main menu pulldown list (Switcher)*

The second menu, called "Options", gives direct access to any of the screens in the user interface.

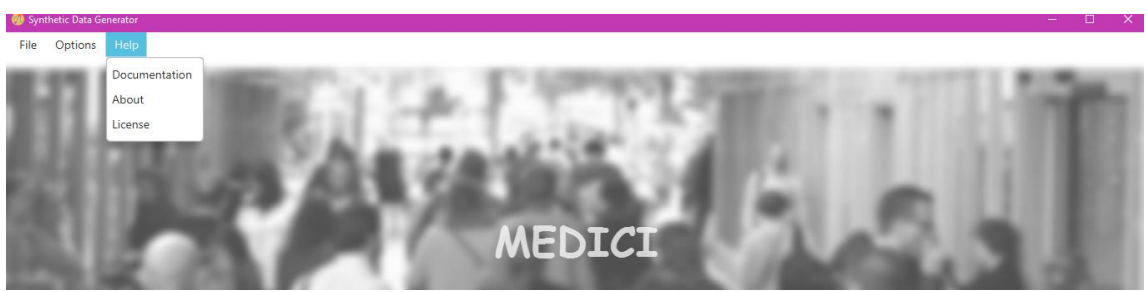

*Fig. 2c Main menu pulldown list (Help)*

Lastly, the "Help" menu has the following options: "Documentation" opens the user manual of the program as a PDF file; "About" opens a pop-up window with the credits which show the authors of the program: "License" opens a pop-up window showing a summary of the software license with a button to see the full license details. The software is licensed under the GNU General Public License 3.0.

**A.2 Home Page**

The first page is shown when the programs starts up (Fig. 3), and just shows a "go" button which then navigates to the first tab, "Generate Graph". Also, the pull down menus are available, as described before.

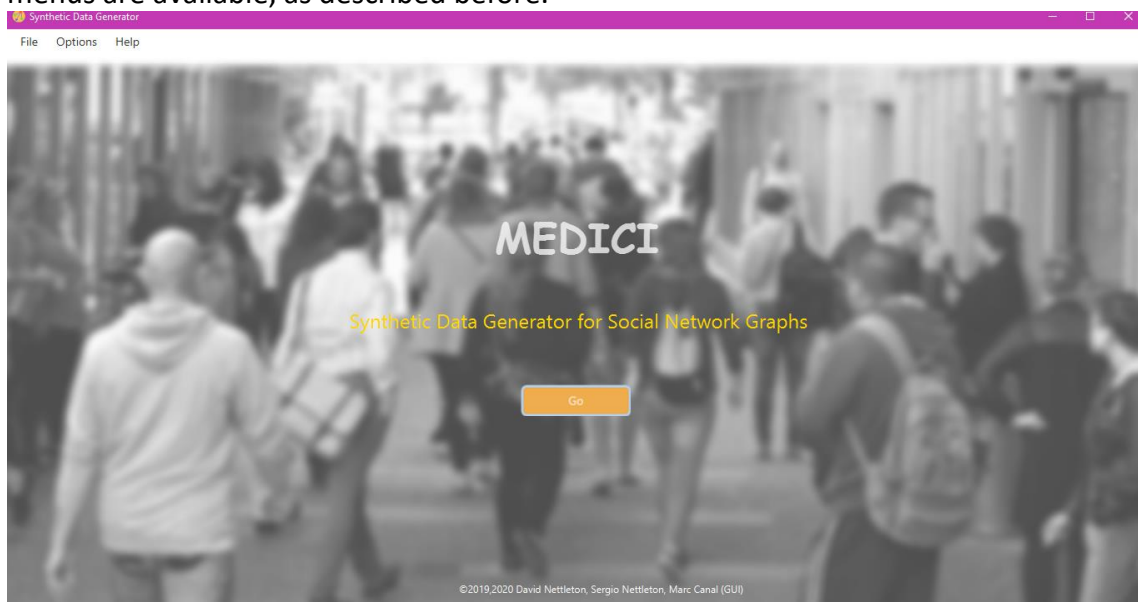

*Fig.3  Home Page*

**A.3 Generate Graph and identify Communities**

Figure 4 shows the first screen of the program "workflow" (going left to right) which allows the user create the graph from scratch using the RMAT graph generator, and then identify the communities using the Louvain community detection algorithm. In the case of RMAT, it is recommended to assign the number of nodes and edges, leaving parameters a to d as default. However, experts may try varying these to evaluate the resulting graphs in terms of community structure. Essentially, parameters a and d define the communities whereas b and c define the links between communities. In the case of Louvain, the algorithm has been customized to find exactly 10 communities, and a "quality" value is shown after processing of the assigned communities. The closer the quality value is to 1, the better the quality.

Note that in the upper part of the screen (Figure 4) is the navegation bar with different tabs common to all workflow screens. There is one tab for each main workflow step, allowing direct access, with the current one being highlighted.

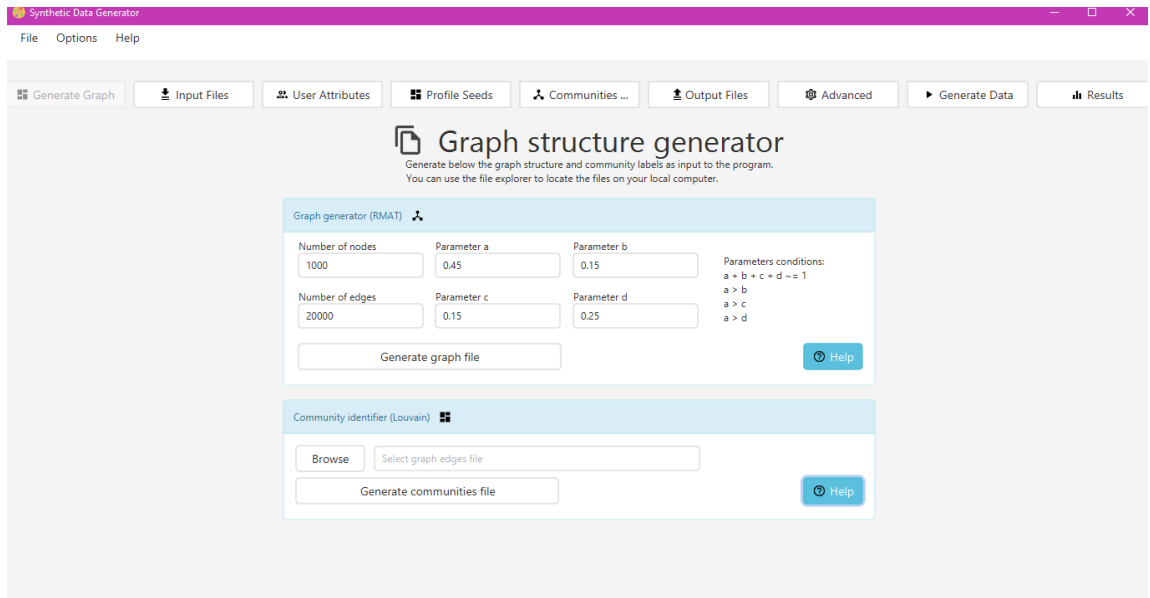

*Fig. 4 Graph Generator Page*

The first file generated is the graph structure, which is a simple list of node pair connections in "txt" format. For example:

**1kby30k.csv**

| 12 | 77 |
|---|---|
| 99 | 697 |
| 843 | 847 |
| 316 | 66 |
| 845 | 331 |
| 557 | 278 |
| 734 | 736 |
| 750 | 752 |
| 166 | 175 |
| 507 | 517 |
| 694 | 688 |
| 328 | 328 |

where in the first line 12;77 defines two nodes (or "users") 12 and 77 and the link between them.

The second input file contains the community label assignments for each node:

**1kby30kcommunities.csv**

| 0 | 0 |
|---|---|
| 1 | 1 |
| 2 | 2 |
| 3 | 1 |
| 4 | 0 |
| 5 | 0 |
| 6 | 1 |
| 7 | 2 |

where in the first line *0;0* indicates that node 0 is assigned to community 0, the second line *1;1* indicates that node 1 is assigned to community 1, and so on.

Note that the default files are in "./resources/Default_files" and if the user generates new graph and community files they will appear in the folder "./resources/Input_files".

**A.4 Input File Settings**

Figure 5 shows the second screen of the program "workflow" which allows the user to select the input files for the graph and the communities, which will be used later by the "generate data" process. The default files are in the system folder "./resources/Default_files". file names will be automatically assigned (in a default system folder which is shown). If the user generates new graph and community files using the RMAT and Louvain algorithms, they will appear in the folder "./resources/Input_files". Also, the user can place his/her own graph and community files in the folder "./resources/Input_files".

The red button allows the user to reset the file names and paths to default, the light blue button displays a popup window with additional information and the green button saves the current assignment and go to the next step in the workflow.

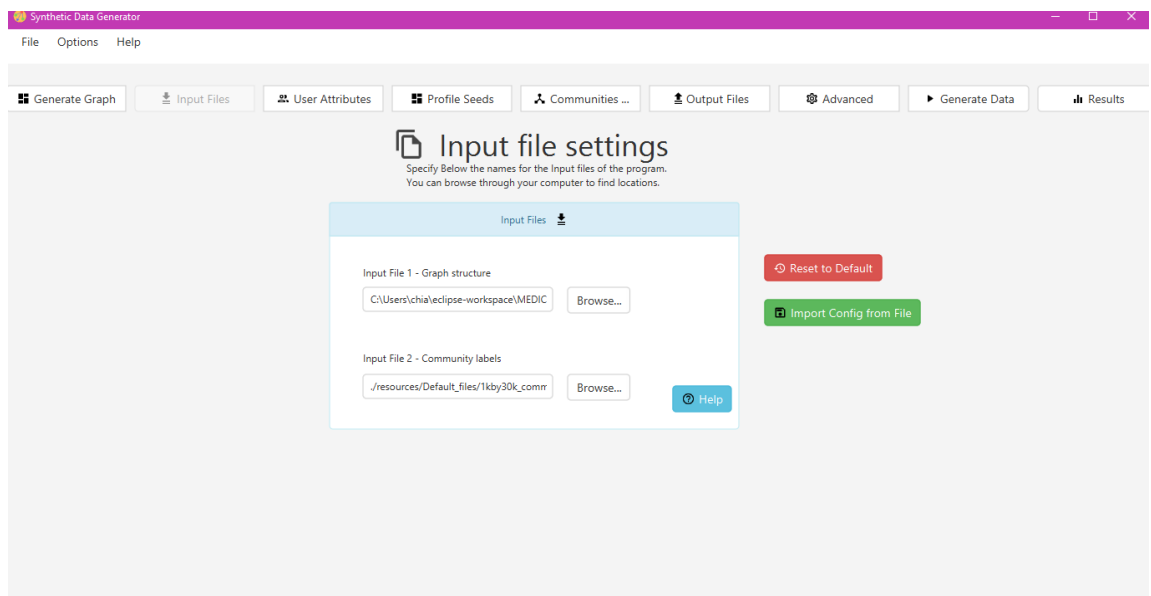

*Fig. 5 Input File Settings Page*

**A.5 User Attributes**

Figure 6 shows the "User Attributes" page which allows the user to visualize the attributes which characterize the individuals in the social network, and modify the relative proportions of each attribute-value. Each attribute is presented as a record with name, description and a list of values with their respective frequencies. For example, on the left, attribute "Age" has seven ranges, the first being 18-25 years with proportion 0.25 and the second 26-35 years with proportion 0.25. In contrast, age range 76-85 years has the relatively smallest proportion of 0.08. Thus the distribution defined for "age" has a strong bias towards younger people.

On clicking "edit" a window will appear to edit the attribute (see next Section A6 and Fig. 7). Note the sum of the proportions for the values of given attribute must sum to 1.0.

In terms of the data generation/propagation, these percentage distributions and categories are used when assigning neighbor and non-neighbors of the seed (profile) nodes, which has a probabilistic nature.

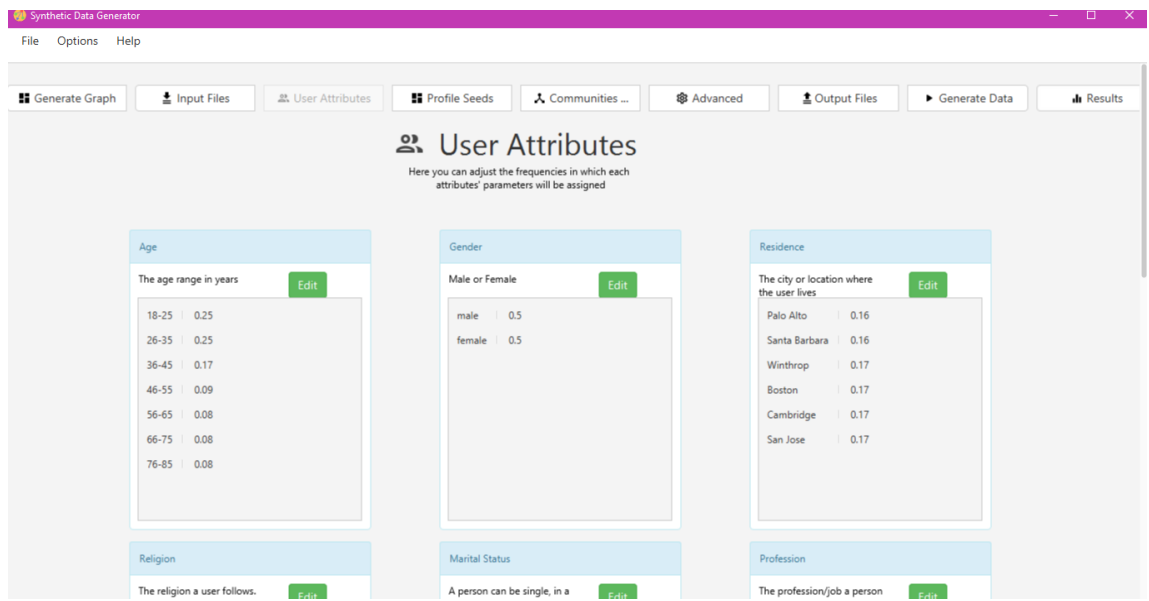

*Fig. 6 User Attributes Page*

**A.6 Attribute Edit Dialog**

Fig. 7 shows the window which allows the user to edit the attribute-value proportions. The attribute record shows the attribute name and description, which is followed by a list of possible values for the attribute and the relative frequency/proportion for each values.

For example, in Fig. 7, the attribute "gender" has been assigned two possible values, "male" and "female", which currently have assigned equal relative proportions of 50% each. If a modification is made, the green "ok" button will save the changes or click the red "cancel" button to quit without saving.

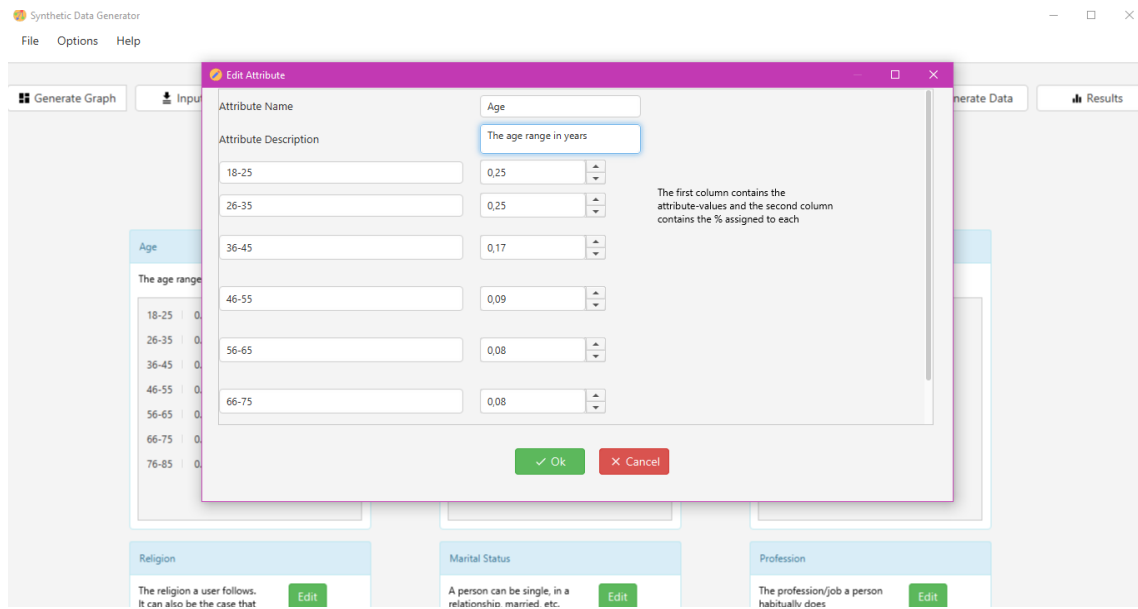

*Fig. 7 Attribute Edit Dialog*

**A.7 Profiles Seeds Page**

Figure 8 shows the "Profile Seed Settings" page. In this screen, different user profiles can be configured. One profile is later assigned to each community (see Section A8), which are fixed to 10 profiles/communities. For each profile there is a pulldown list for each attribute, with the assignable values. There is also a (red) button on the right to return all the fields to their default values. For example, in Fig. 8, profile 0, attribute "age" is set to the range "36-45" and the "marital status" attribute is set to "Married".

The user can customize a profile by choosing an "attribute-value" for each of the attributes which describe a "seed" (or prototype) individual of the social network. The system allows for up to 10 profiles to be created, one for each community. When the data is generated the Medici algorithm probabilistically assigns the seeds "neighbors" with profiles which are similar to the seed, with a controlled amount of random noise.

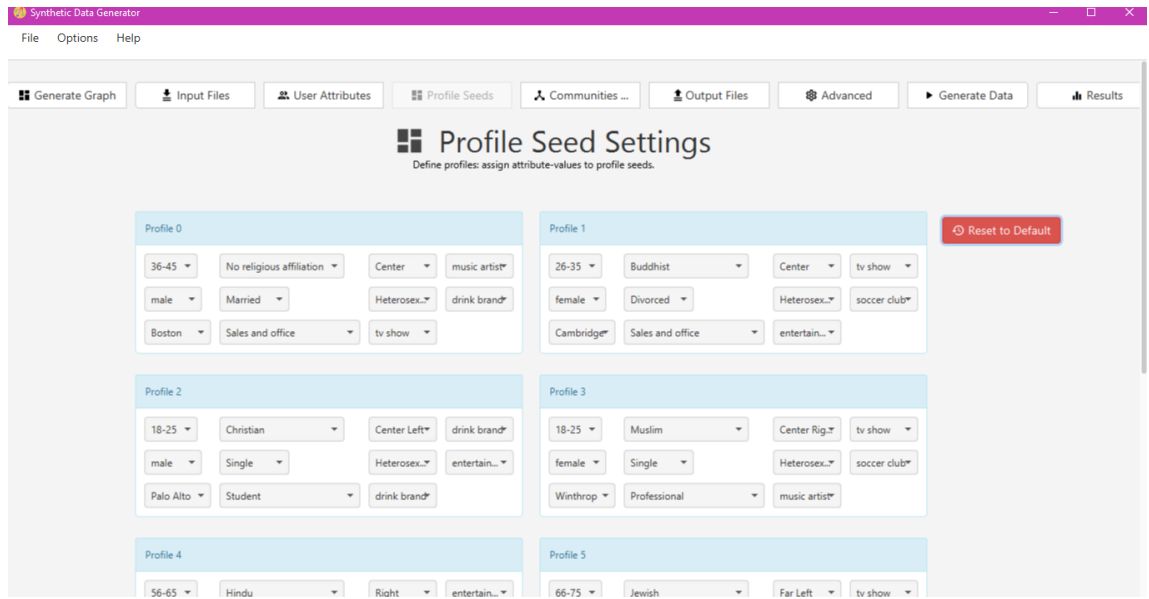

*Fig. 8 Profile Seed Settings Page*

**A.8 Communities and Profiles**

Once the profile seeds have been defined the next step is to assign them to the communities. The communities have relative proportions depending on how many nodes they have assigned, which is defined when the graph is generated and the community labels assigned. Hence, if you wish to have a majority of users with a given profile $P^n$ in the overall graph, you find the biggest community and assign the profile $P^n$ to it. Contrastly, if you want a just a few users to have a given profile, you find a community with a small proportion and assign the profile to it.

Profile to Community assignment is accessed using the screen shown in Fig. 9. This screen has a button which opens the window to assign the profiles to communities, and includes informational text explaining the functionality.

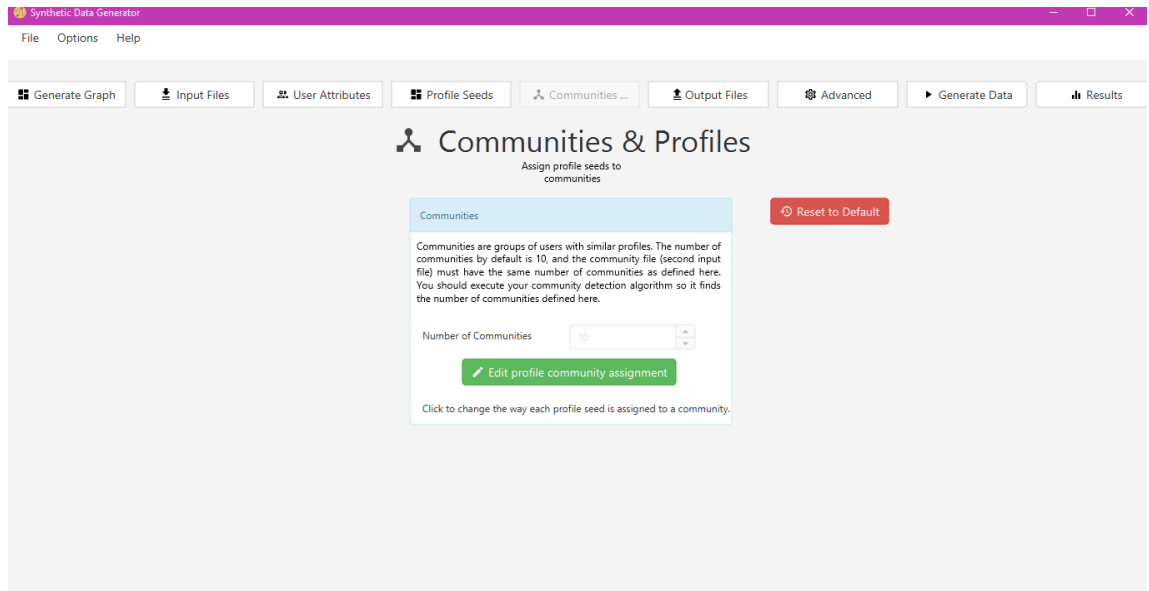

Fig. 9 Communities & Profiles page

Click on the green button labelled "Edit profile community assignment" (Fig. 9) to open the screen (Fig. 10) which allows the user to see the frequency/proportion for each community and assign the profiles. The total (sum for all profiles) is shown which must sum to 100% (this is calculated automatically from the graph). For example, profile 0 has a proportion of 21.6%, and profile 7 has just 0.5%. It can be seen that profiles 0,1,2 and 5 have the highest proportions and profiles 7 and 9 have the lowest.

Click on the green button labelled "Edit profile community assignment" (Fig. 9 right) to open the screen which allows the user to assign the profiles to communities (Fig. 10b). In this screen there is a list of communities and for each community a pulldown list where there user can assign the profile. It can be seen that profile 0 has been assigned to community 0 and profile 6 to community 8. The graph generator and community labeller produce a distribution of proportions similar to a "long tail", within the limitations offered by 10 values, thus having a similar community size distribution to real world social networks.

*Note that the user can provide his/her own graph and community files, instead of generating them in Medici, but then the responsibility for their correctness relies on the user.*

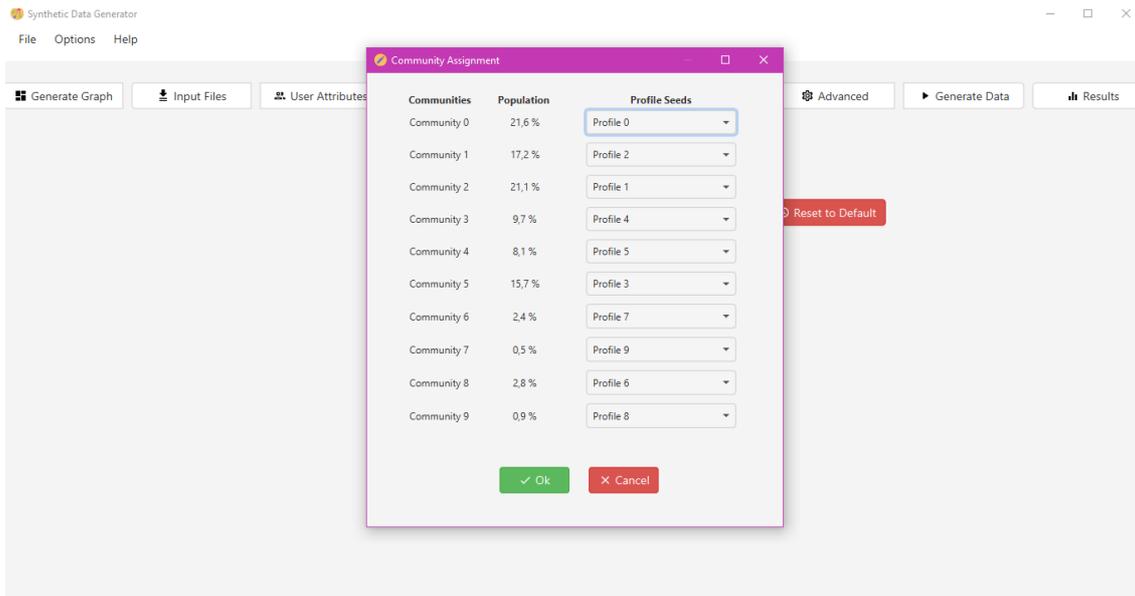

*Fig. 10 Assignment of Profiles to Communities*

**A.9 Output Files**

Figure 11 shows the screen to assign the output files. It is similar to the input file screen (Fig. 5), with a similar functionality. It allows the user to specify where to save and what to call the files which contain the results of generating the social network user data. The default folder is "./resources/Output_files/". The user can also save the current configuration (attribute and profile assignments, etc.) by clicking the green button "Export Config to File" on the left. The red button restored the default system values.

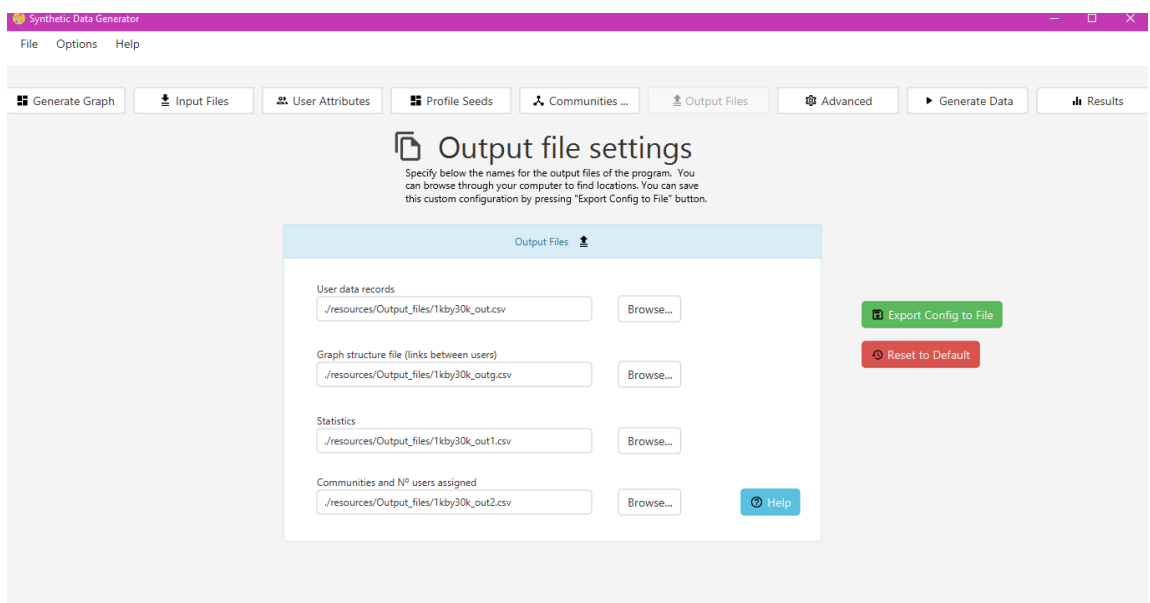

*Fig. 11 Output File Settings*

The format of each output file is as follows:

**User data records** - default name "1kby30k_out.csv"

| user | age | gender | residence | religion | maritalstatus | profession | politicalorientation | sexualorientation | numfriends | like1 | like2 | like3 | classvalue | auth | community |
|---|---|---|---|---|---|---|---|---|---|---|---|---|---|---|---|
| 999 | 26-35 | female | Winthrop | Hindu | Divorced | Professional | Left | Heterosexual | LOW | entertainment | entertainment | music artist | NO | 0.026178 | 5 |
| 998 | 26-35 | female | Cambridge | Buddhist | Single | Service | Center Right | Heterosexual | LOW | entertainment | entertainment | entertainment | YES | 0.041885 | 6 |
| 997 | 26-35 | female | Boston | Jewish | Married | Service | Left | Heterosexual | LOW | music artist | music artist | entertainment | NO | 0.031414 | 3 |
| 996 | 18-25 | female | Winthrop | Muslim | Married | Professional | Center Right | Heterosexual | LOW | music artist | music artist | drink brand | NO | 0.068063 | 5 |
| 995 | 66-75 | female | San Jose | Jewish | Married | Production | Far Left | Heterosexual | LOW | entertainment | tv show | entertainment | NO | 0.036649 | 4 |
| 994 | 18-25 | female | Winthrop | Muslim | Single | Production | Center Right | Heterosexual | LOW | drink brand | tv show | drink brand | NO | 0.073298 | 5 |
| 993 | 26-35 | female | Cambridge | Buddhist | Widowed | Sales and o | Center Left | Homosexual | LOW | entertainment | tv show | soccer club | YES | 0.031414 | 2 |
| 992 | 26-35 | female | Cambridge | Christian | Widowed | Sales and o | Left | Heterosexual | LOW | music artist | music artist | entertainment | YES | 0.115183 | 8 |
| 991 | 56-65 | female | Palo Alto | Jewish | Married | Production | Left | Bisexual | LOW | music artist | music artist | entertainment | NO | 0.041885 | 4 |
| 990 | 56-65 | male | Santa Barba | Hindu | Widowed | Natural res | Left | Heterosexual | LOW | tv show | entertainment | drink brand | NO | 0.120419 | 9 |
| 989 | 56-65 | male | Boston | Buddhist | Single | Sales and o | Center Left | Heterosexual | LOW | music artist | music artist | drink brand | NO | 0.08377 | 0 |
| 988 | 26-35 | female | Cambridge | Hindu | Married | Student | Left | Heterosexual | LOW | drink brand | tv show | drink brand | YES | 0.078534 | 2 |
| 987 | 18-25 | female | Winthrop | Muslim | Single | Professional | Center Right | Heterosexual | LOW | music artist | tv show | soccer club | NO | 0.068063 | 5 |
| 986 | 56-65 | male | San Jose | Hindu | Single | Sales and o | Center Right | Homosexual | LOW | tv show | drink brand | soccer club | NO | 0.125654 | 3 |
| 985 | 26-35 | female | Cambridge | Buddhist | Married | Professional | Left | Heterosexual | LOW | entertainment | tv show | soccer club | YES | 0.115183 | 2 |
| 984 | 26-35 | female | San Jose | Sikh | Divorced | Student | Center | Heterosexual | LOW | drink brand | tv show | drink brand | NO | 0.198953 | 2 |
| 983 | 56-65 | female | Cambridge | Hindu | Divorced | Production | Far Left | Heterosexual | LOW | entertainment | entertainment | music artist | NO | 0.057592 | 9 |
| 982 | 26-35 | female | Cambridge | Buddhist | Single | Sales and o | Center | Heterosexual | LOW | entertainment | tv show | soccer club | YES | 0.115183 | 2 |
| 981 | 56-65 | female | San Jose | Hindu | Single | Natural res | Right | Heterosexual | LOW | music artist | entertainment | entertainment | NO | 0.089005 | 3 |
| 980 | 66-75 | female | San Jose | Christian | Single | Production | Far Left | Bisexual | LOW | tv show | drink brand | soccer club | NO | 0.141361 | 4 |
| 979 | 26-35 | female | Palo Alto | Buddhist | Single | Manager | Center Left | Homosexual | LOW | music artist | music artist | entertainment | NO | 0.078534 | 0 |
| 978 | 18-25 | female | Winthrop | Muslim | Single | Professional | Center Right | Heterosexual | LOW | music artist | tv show | soccer club | NO | 0.136126 | 5 |
| 977 | 46-55 | male | Boston | Hindu | Married | Manager | Center | Heterosexual | LOW | entertainment | entertainment | music artist | NO | 0.08377 | 0 |
| 976 | 18-25 | female | Winthrop | Muslim | Single | Professional | Center Right | Heterosexual | LOW | music artist | tv show | soccer club | NO | 0.136126 | 5 |
| 975 | 18-25 | female | San Jose | Buddhist | Widowed | Student | Center | Heterosexual | LOW | drink brand | tv show | drink brand | NO | 0.052356 | 1 |
| 974 | 26-35 | female | Cambridge | Hindu | Married | Student | Left | Heterosexual | LOW | entertainment | entertainment | music artist | YES | 0.089005 | 2 |

Contains the social network user information, one row per individual, where the first row is the header and each consecutive row consists of the user (node) id followed by the value for each attribute for this individual. For example, user id 999 has "gender" attribute assigned with value "female" and has "residence" attribute assigned with value "Winthrop", etc.

**Graph structure file (links between users)** - default name "1kby30k_outg.csv"

| user | userf | linkweight |
|---|---|---|
| 999 | 487 | 0.57 |
| 999 | 658 | 0.54 |
| 999 | 428 | 0.62 |
| 999 | 731 | 0.30 |
| 999 | 57 | 0.60 |
| 998 | 448 | 0.47 |
| 998 | 791 | 0.89 |
| 998 | 937 | 0.70 |
| 998 | 299 | 0.55 |
| 998 | 656 | 0.56 |
| 998 | 844 | 0.50 |

Contains the graph information, one row per link. For example, in the first row user 999 has a link to user 487 with a link strength of 0.57, and the second row shows that user 999 also has a link to user 658 with strength 0.54, etc.

**User ids** – default name "1kby30k _out1.csv"

| community | attribute | value | frequency |
|---|---|---|---|
| ALL | AGE | 18-25 | 275 |
| ALL | AGE | 26-35 | 299 |
| ALL | AGE | 36-45 | 182 |
| ALL | AGE | 46-55 | 46 |
| ALL | AGE | 56-65 | 106 |
| ALL | AGE | 66-75 | 64 |
| ALL | AGE | 76-85 | 28 |
| ALL | GENDER | Male | 329 |
| ALL | GENDER | Female | 671 |
| ALL | RESIDENCE | PaloAlto | 141 |
| ALL | RESIDENCE | SantaBarbara | 141 |
| ALL | RESIDENCE | Winthrop | 195 |
| ALL | RESIDENCE | Boston | 206 |
| ALL | RESIDENCE | Cambridge | 198 |
| ALL | RESIDENCE | SanJose | 119 |
| ALL | RELIGION | Buddhist | 214 |
| ALL | RELIGION | Christian | 211 |
| ALL | RELIGION | Hindu | 287 |
| ALL | RELIGION | Jewish | 87 |
| ALL | RELIGION | Muslim | 157 |
| ALL | RELIGION | Sikh | 20 |

Contains the statistics of the data generated for the users, for ALL the graph and then for each community in turn. For example, the first row shows that for ALL the graph, attribute-value "age 18-25" has a frequency of 275.

**Communities and Nº users assigned** – default name "1kby30k_out2.csv"

| | |
|---|---|
| 9 | 9 |
| 8 | 28 |
| 7 | 5 |
| 6 | 24 |
| 5 | 157 |
| 4 | 81 |
| 3 | 97 |
| 2 | 211 |
| 1 | 172 |
| 0 | 216 |

Contains the summary information for each community. For example, the first row indicates that 9 users are assigned to community 9 and the last row indicates that 216 users are assigned to community 0.

**A.10 Advanced Settings**

Figure 12 shows the "Advanced Settings" screen, which displays the options for changing the advanced features of the program. This is only for expert users, and the "thresholds" option is disabled in the current version of Medici. There are three options: "Profile seeds", "Thresholds" and "Randomness", which have a brief explanatory text on screen and additional information is available by pressing the light blue "help" button. In the majority of situations the program should function optimally with the default settings of "seeds percentage=11" and "random assignment ratio=low".

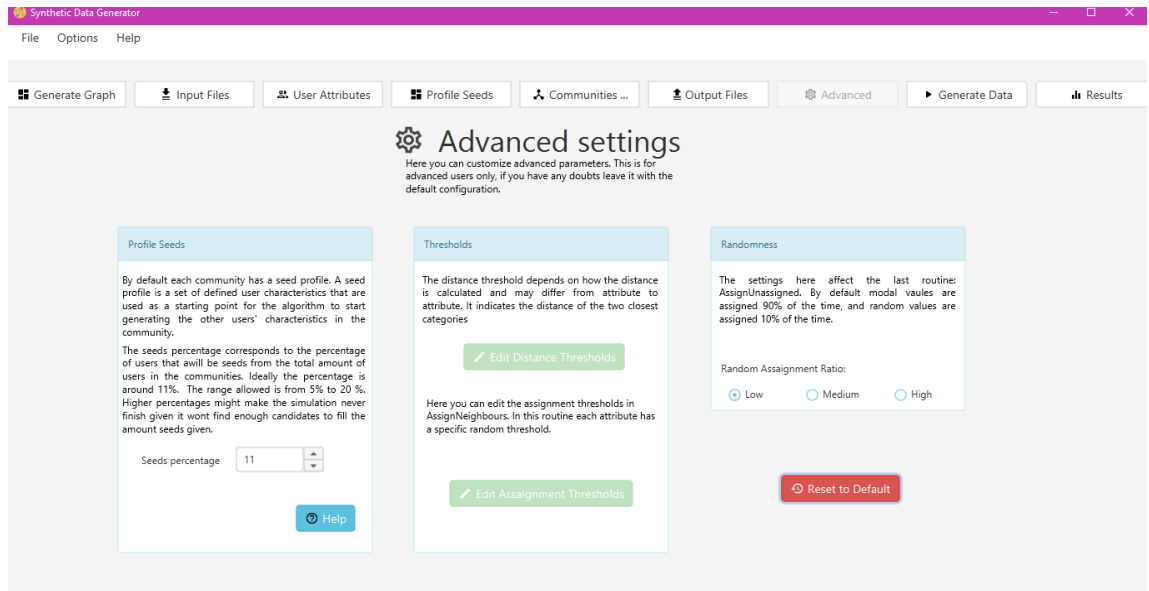

*Fig. 12 Advanced Settings Screen*

**A.11 Generate Data Screen**

Figure 13a shows the "Generate Data Screen" where the user finally executes the data generation based on all the attributes, profiles and community assignments which have been assigned previously. The main button "Generate Data" executes the back-end Medici algorithm and while it is running a loading bar appears to show progress (Fig 13b). There is also a red button to cancel the process. These two elements disappear once the data is generated and a pop up window will appear to indicate to the user that the generation has terminated, with a count of how many nodes have been assigned (seeds + neighbours), which in Figure 13b is shown as 786, of the 1000 total nodes in the graph. This means that the remaining 214 nodes will be assigned probabilistically (due to closeness to an assigned node) or randomly.

If any errors have occurred a pop up will appear with a red cross to inform of any problems.

If the process has terminated correctly, the user can now click on the "results" tab to see the statistics of the resulting data and access the output files.

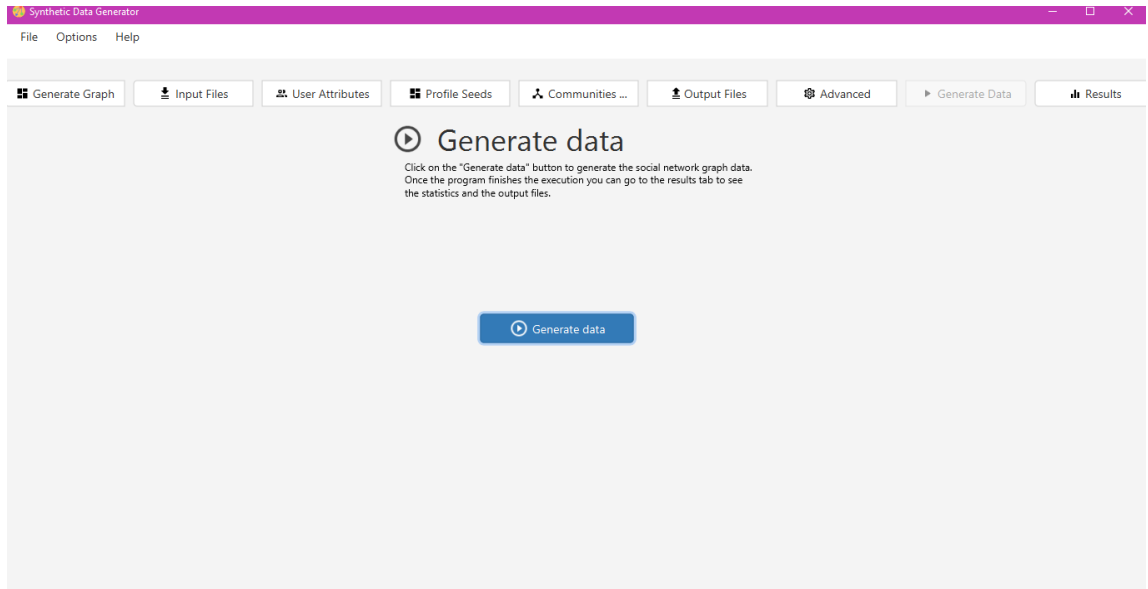

Fig. 13a Generate Data Screen

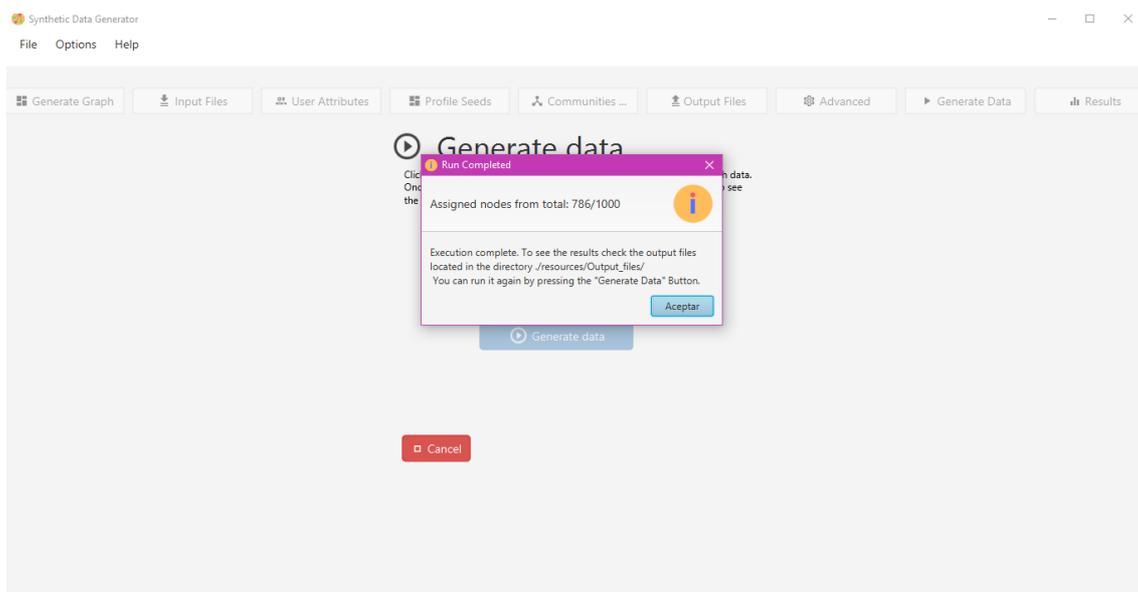

Fig. 13b Run Screen

**A.12 Results Screen**

Figure 14(a to c) show the results screen with the statistics for the data which has been generated Fig. 14a shows the statistics for the whole graph (all communities), and Figs. 14b and 14c show the statistics just for community 2 (profile1). On the top left of the screen there is an option to select all the data or a specific community. On the top right there are options to open the different results files.

Below there is the statistics panel which can be filtered using the pulldown list on the top right. This allows the user to see the global statistics or filter on a given community. The statistics show a graphic with the proportion for each community and corresponding profile seed (pie chart on left), then for each attribute a graphic with the percentage of assigned values (e.g. age and gender to the right).

Each attribute is shown as a frequency bar graph with a bar for the profile percentage assigned by the user to the "seeds" and another bar with the frequency for the same attribute in that community.

Here it is important to note that the two percentages will not necessarily be the same, because the algorithm, as described previously, assigns the "pure" profiles only to the seeds in a given community. The neighbors of each seed will be similar but not necessarily the same (e.g. 30% could be different), due to how the propagation works, with a predefined stochastic (random) behavior which guarantees diversity. Furthermore, the remaining nodes (not neighbors of any seed) will have a further diversity. Hence it is normal that the profile frequency and the community frequency will be different most of the time. However, overall, the community should have a distinct identifiable similarity to the assigned profile attribute distributions, which distinguishes it from other communities with different profile assignments.

So in Figure 14a it can be seen that overall (general trends in the graphs) there are three main age groups, 18-25, 26-35 and 36-45, the gender has a bias towards female, and so on.

Then in Figure 14b, the statistics for community 2 (profile1) show distinct differences to the general trends. For example, age is predominantly 26-35, and gender has a much greater proportion of female (with respect to whole graph), which corresponds to the profile1 proportion assigned to community 2. Finally, Figure 14c shows the tabular summary of the profile seed assignments vs the real data generated for a community. Again note that the real percentage also depends on the diversity factor, thus the community should have a distinct bias towards the assigned profile1 (when compared with other profiles) but it will not be identical.

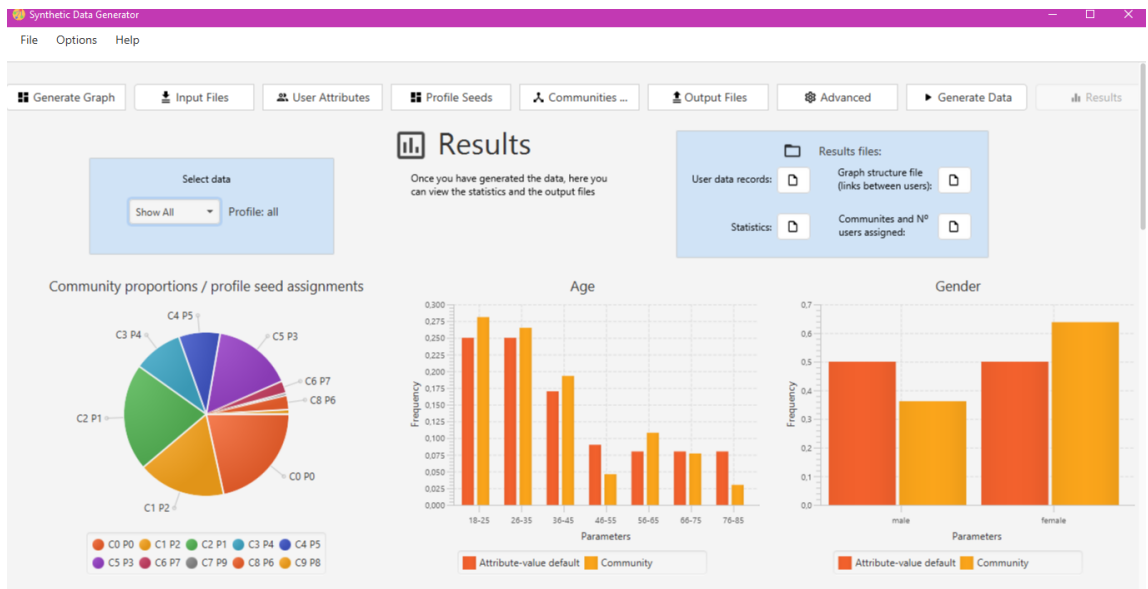

*Fig. 14a Statistics Page (all communities)*

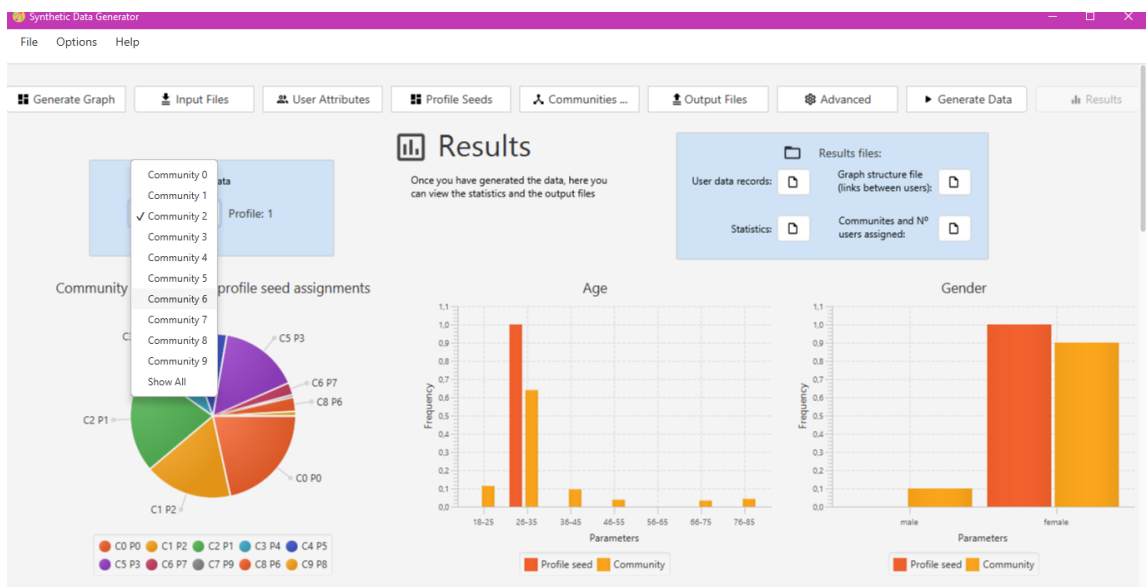

*Fig. 14b Statistics Page (community 2)*

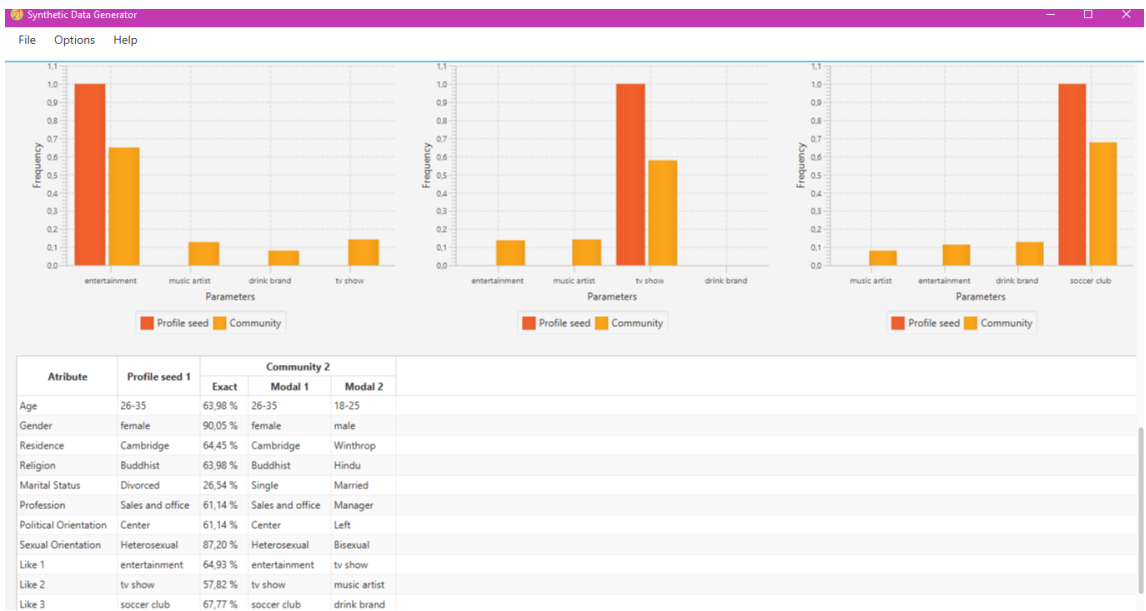

Fig. 14c Statistics Page (community 2)

### Annex B. Pseudo code of data generator/propagator

As described in Section 3 of the paper, the data generator/propagator has four main steps: **(i)** For each community, choose which nodes will be seeds; **(ii)** For each community, assign corresponding prototype profile and then use it to assign data to seeds (profile x => community x); **(iii)** Assign data to neighbors of each seed in function of seed profile (i.e. neighbors tend to be similar to seeds). This will have a similarity component (to neighbors with data assigned) and a random component (to promote a degree of diversity); **(iv)** Assign data to remaining nodes (those which still have no data assigned from steps (ii) and (iii). This will have a similarity component (to neighbors with data assigned) and a random component (to promote a degree of diversity).

The following psuedo code procedures correspond to the above steps as follows: **(i)** "Seed Assigner"; **(ii)** and **(iii)** "Synthetic Data Generator"; **(iv)** "Assign Unassigned Vertices in Community".

**Procedure Seed Assigner**
*Input:* graph G, δ=0.05
*Output:* seed set S
1. **Assign** *Seeds*
2.     Let $\phi$ be the average degree of all vertices V in G.
3.     $\sigma = |V| / \phi$
4.     **While** S does not comply **do**
5.      **While** S is not topologically optimal **do**
6.       **Choose** a set S of $\sigma$ seeds, in which:
          Each s ∈ S is at least at distance 3 from any other seed.
          The number of seeds assigned is equal to $\sigma$.
        **Check** optimality of S
7.      **End do**
8.      **Check** that distributions of key metrics and attribute-values of s ∈ S are within margin δ of corresponding distributions of v ∈ G
9.       Degree distribution, clustering coefficient, distributions of all attributes.
10.      **Assign** compliance of S
11.     **End do**

**Procedure Synthetic Data Generator**
*Input:* Number of vertices and edges, $p$ =degree of data diversity
*Output:* graph G
1. *RMat*
2.     **For** |V| vertices and |E| edges **generate** an OSN-like topology.
3. *Communities*
4.     **Calculate** communities using Leuven method and assign community tag to each vertex.
5. *Authorities and dense sub-graphs*
6.     **Calculate** top authorities $A_c$ and dense sub-graphs $D_c$,
      $A_c$, $D_c$ ∈ $AD_c$ in each community $c$ using HITS algorithm and clustering coefficient metric
7.     Each vertex $ad_c$ must be at a distance ≥ 3 from any other authority or dense sub-graph in community c
8. *Assign data to AD's in each community*
9. **For each** community c do
10. **For each** vertex $ad_c$ ∈ $AD_c$ **do**
11.     **For each** edge e connected to $ad_c$, assign a random

|       | weight between 0 and 10. |
|-------|---|
| 12.   | **Assign** attribute-values to $ad_c$ |
| 13.   | **Assign** *attribute-values to neighbors of* $ad_c$ |
| 14.   | Let $Nad_c$ be the set of neighbors of $ad_c$ |
| 15.   | **For each** $n \in Nad_c$ **do** |
| 16.   |     **For each** attribute a of n **do** |
| 17.   |         **For each** value v of attribute n **do** |
| 18.   |             **Probability of assignment of {n, a, v} = p** |
| 19.   |             Assign {a, v} of $ad_c$ to neighbor *n* with probability 1- *p* |
| 20.   |         **End do** |
| 21.   |     **End do** |
| 22.   | **End do** |
| 23.   | Let $NA_c$ be the set of vertices in c with data assigned |
| 24.   | *Assign data to remaining nodes in each community* |
| 25.   | **Call** **Assign Unassigned Vertices in Community**($NA_c$ , c) |
| 26.   | **End do**  // for each community |

**Procedure Assign Unassigned Vertices in Community**

*Input:* $NA_c$ , the set of vertices in c with data assigned; c, the current community id
*Output:* assigned vertices in community c

| 1. | **For each** n in c $\notin NA_c$ **do** |
|---|---|
| 2. |     **For each** attribute a of n **do** |
| 3. |         **For each** value v of attribute n **do** |
| 4. |             **Calculate** average or modal value of corresponding attribute-value of neighbors of n as {n', a', v'} |
| 5. |             **Probability of assignment of {n', a', v'} = p** |
| 6. |             Assign {a', v'} to *n* with probability 1- *p* <br>     *Otherwise* <br>       Assign random neighbor {a'', v''} to n if at least one non-null value <br>     *Otherwise* <br>       Assign random values {a''', v'''} |
| 7. |     **End do** |
| 8. | **End do** |